\documentclass{article}

\usepackage{arxiv}

\usepackage[utf8]{inputenc} % allow utf-8 input
\usepackage[T1]{fontenc}    % use 8-bit T1 fonts
\usepackage{hyperref}       % hyperlinks
\usepackage{url}            % simple URL typesetting
\usepackage{booktabs}       % professional-quality tables
\usepackage{amsfonts}       % blackboard math symbols
\usepackage{comment}
\usepackage{rotating}
\usepackage{caption} % for '\ContinuedFloat' directive
\usepackage{tabularx}

\usepackage{fancyhdr}
\usepackage{graphicx}
\usepackage{authblk}
\usepackage{natbib}

\newcommand{\akatsuki}{\textit{Akatsuki}\,}
\newcommand{\mps}{$\,\textrm{m}\cdot\textrm{s}^{-1}\,$}

\title{Amateur Observers Witness the Return of Venus' Cloud Discontinuity}

\author[1]{E. Kardasis\thanks{Corresponding author: E. Kardasis (astromanos2002@yahoo.gr).}}
\author[2]{J. Peralta}
\author[1,3,4]{G. Maravelias}
\author[5]{M. Imai}
\author[6]{A. Wesley}
\author[7]{T. Olivetti}
\author[8]{Y. Naryzhniy}
\author[9,10]{L. Morrone}
\author[11]{A. Gallardo}
\author[12]{G. Calapai}
\author[13]{J. Camarena}
\author[14]{P. Casquinha}
\author[15]{D. Kananovich}
\author[10,16]{N. MacNeill}
\author[17]{C. Viladrich}
\author[1]{A. Takoudi}

\affil[1]{Hellenic Amateur Astronomy Association, Alopekis 42, 10676,Athens, Greece.}
\affil[2]{Departamento de FAMN, Facultad de F\'{i}sica, Universidad de Sevilla, Sevilla 41012, Spain.}
\affil[3]{IAASARS, National Observatory of Athens,Lofos Nymfon, Thission, 118 10,Athens, Greece.}
\affil[4]{Institute of Astrophysics, FORTH, Heraklion, Greece.}
\affil[5]{Faculty of Science, Kyoto Sangyo University, Japan.}
\affil[6]{Astronomical Society of Australia, Rubyvale, QLD 4702, Australia.}
\affil[7]{Union of Italian Amateur Astronomers, Via Lazio, 14-00040 Rocca di Papa, RM, Italy.}
\affil[8]{Kagarlyk, Kiev Region, Ukraine.}
\affil[9]{AstroCampania Association,Agerola, Italy.}
\affil[10]{British Astronomical Association, Burlington House, Piccadilly, London W1J 0DU, UK.}
\affil[11]{Asociaci\'{o}n Astron\'{o}mica del Campo de Gibraltar ``Luz Cero'', Spain.}
\affil[12]{Private Astronomical Observatory in Massa S. Giorgio, Messina, Italy.}
\affil[13]{Agrupaci\'{o}n Astron\'{o}mica de la Safor, Valencia, Spain.}
\affil[14]{Portuguese Association of Amateur Astronomers, Portugal.}
\affil[15]{Faculty of Chemistry,Tallinn University of Technology, Tallinn, Estonia.}
\affil[16]{ Astronomy Society of NSW, PO Box 870 Epping NSW 1710, Australia.}
\affil[17]{Astroqueyras, France.}

\begin{document}
\maketitle

\begin{abstract}
Firstly identified in images from JAXA's orbiter \akatsuki, the cloud discontinuity of Venus is a planetary-scale phenomenon known to be recurrent since, at least, the 1980s. Interpreted as a new type of Kelvin wave, this disruption is associated to dramatic changes in the clouds' opacity and distribution of aerosols, and it may constitute a critical piece for our understanding of the thermal balance and atmospheric circulation of Venus. Here, we report its reappearance on the dayside middle clouds four years after its last detection with \akatsuki/IR1, and for the first time, we characterize its main properties using exclusively near-infrared images from amateur observations. In agreement with previous reports, the discontinuity exhibited temporal variations in its zonal speed, orientation, length, and its effect over the clouds' albedo during the 2019/2020 eastern elongation. Finally, a comparison with simultaneous observations by \akatsuki UVI and LIR confirmed that the discontinuity is not visible on the upper clouds' albedo or thermal emission, while zonal speeds are slower than winds at the clouds' top and faster than at the middle clouds, evidencing that this Kelvin wave might be transporting momentum up to upper clouds.
\end{abstract}

% keywords can be removed

\section{Introduction}\label{sec:intro}
Venus is permanently covered by a thick stratified atmosphere. Within altitudes \mbox{$\sim$48--70 km} above the surface, we find clouds of sulfuric acid \citep{Titov2018}. These are divided into upper, middle, and lower clouds. At the top (56.5--70 km above the surface), the clouds are dominated by a retrograde superrotation that reaches speeds of $\sim$100\mps {at the clouds' top of lower latitudes}, which is 60 times faster than the planetary rotation \citep{Sanchez-Lavega2017}. This is the layer that has been observed continuously for almost a century with violet and ultraviolet (UV) filters~\citep{Wright1927,Ross1928}, whose dark/bright features are caused by known (SO$_2$) and unknown absorbers \citep{Titov2018,Perez-Hoyos2018}. Under this layer, the middle clouds (50.5--56.5 km above the surface) are observed on the dayside at visible and near-infrared (NIR) wavelengths up to 1 $\mu$m \citep{Peralta2019GRL,Titov2018}. The lower clouds (47.5--50.5 km) can be observed as silhouettes partially blocking the deeper thermal emission on the nightside {mainly at the spectral windows 1.74 $\mu$m and 2.32 $\mu$m} \citep{Limaye2018EPS,McGouldrick2012, Titov2018}. The middle and lower clouds contribute to the greenhouse effect and the radiative energy balance \citep{Peralta2020}. The morphology and motions of the clouds differ at each of the three layers \citep{Horinouchi2018,Limaye2018EPS,Peralta2019GRL,Sanchez-Lavega2017, Titov2018}.

Multiple observations have provided evidence of planetary-scale waves in this stratified atmosphere, and these waves are suspected to play a critical role in powering of the superrotation \citep{Sanchez-Lavega2017}. These have been mainly observed at the upper clouds, such as the Y feature~\citep{Peralta2015}, the stationary bow-shaped wave generated at the surface \citep{Fukuhara2017NatGeo}, or the solar tides, which has been shown to maintain the superrotation near the equator \citep{Horinouchi2020}. Recently, \citet{Peralta2020} reported the discovery of a giant discontinuity/disruption at the deeper clouds of Venus, which was shown to be a cyclical phenomenon {limited} between 30$^{\circ}$N and 40$^{\circ}$S, which importantly alters clouds' properties and aerosols \citep{Peralta2020,McGouldrick2021}, propagates faster than the superrotating winds, and is able to keep a coherent shape after several revolutions. Based on numerical simulations, \citet{Peralta2020} interpreted this phenomenon as a new type of planetary-scale wave that may be feeding the superrotation at the upper clouds by bringing momentum from the deeper atmosphere, which is precisely where the Venus atmosphere {accumulates} most of its angular momentum \citep{Schubert1983}. Its absence at the top of the clouds and range of phase speeds also support that this wave should dissipate within the upper clouds \citep{Peralta2020}. \citet{Peralta2020} also showed that the cloud discontinuity (CD henceforth) has been a recurrent feature for decades at the nightside lower clouds, although it is less frequent at the dayside middle clouds where  it was observed for the last time on 27 November 2016 by the camera \akatsuki/IR1, before this instrument and IR2 stopped working indefinitely \citep{Iwagami2018}.

Here, we report the reappearance of the CD on the dayside middle clouds during the 2019/2020 eastern elongation of Venus, using observations from small private telescopes around the world coordinated with JAXA's orbiter \akatsuki \citep{Nakamura2016}. In Section \ref{s:obs+methods}, the outline of the {observations} and the data reduction method are presented. The results of our analysis are provided and discussed in Section \ref{s:results}. Finally, Section \ref{s:summary}, summarizes our work.

\section{Observations and Methods}
\label{s:obs+methods}
\subsection{Venus Images Taken with Ground-Based Telescopes}
To study the CD events, we inspected the amateur observations of Venus hosted in the database of ALPO-Japan 
~{\url{http://alpo-j.sakura.ne.jp/indexE.htm}} (accessed  on  2021) \citep{Sato2018}. We selected and examined 458 NIR and 552 UV observations of the eastern elongation 2019/2020. The NIR observations were obtained with filters in the range {685--1050} nm and the UV observations with filters in the $\sim$320--{410} nm range. Table~\ref{table:NIR-observations} provides details about these NIR {and UV} observations {as well as the Venus phase angle and the nominal best resolution.}   The resolution of the amateur images ranged from 187 to 663 km depending on the {apparent} angular size of Venus {and} the best resolution allowed by the diameter of the telescope  \citep{Sanchez-Lavega2016}. {One of us (EK), on 11 March 2020 (with Venus having a solar elongation of 45.6$^{\circ}$ East, $20.6''$ in diameter, and 58\% of illuminated disk), identified a strong manifestation of a CD and alerted Venus observers and researchers. That event stimulated intense monitoring and led to the current work. We} inspected a period covering from {13 October 2019 to \mbox{25 April 2020}} (Solar Elongation 16.2$^{\circ}$--40.8$^{\circ}$ East, angular size 10.3$''$--35.8$''$, illuminated fraction 0.962--0.295).

% Example of a page in landscape format (with table and table footnote).
%\startlandscape

\begin{sidewaystable}
%\begin{table}[H] 
%%\begin{longtable}

%\small
%\centering
\caption{Summary of NIR {and UV} images of Venus used in this work. We present the date and time of each observation (column 1), the Venus phase angle (column~2), the telescope diameter and type (column 3), the best spatial resolution (column 4), which was calculated theoretically from telescope's properties and apparent disk diameter, the  effective transmission {of the equipment used} (column 5),  the integration time (column 6), which corresponds to the total exposure time for each image, {the observer (column 7), and country (column 8).}}
\label{table:NIR-observations}
\begin{tabularx}{\textwidth}{cccccccc}
\toprule

\textbf{Date and Time (UT)}  & \textbf{Venus Phase} & \textbf{Telescope}  & \textbf{Best Spatial}  & \textbf{Effective} &  \textbf{Integration} & \textbf{Observer}  &
\textbf{Country} \\

\textbf{(YYYY-MM-DDThh:mm.m)} & \textbf{Angle (deg)} &  \textbf{Diameter (m)} & \textbf{Resolution (km)} & \textbf{Transmission (nm)} & \textbf{Time (s)} & & 

%\textbf{Date and Time (UT)} \linebreak \textbf{(YYYY-MM-DDThh:mm.m)} & \textbf{Venus Phase Angle} \linebreak \textbf{(deg)} & \textbf{Telescope Diameter} \linebreak \textbf{(m)} & \textbf{Best Spatial Resolution} \linebreak \textbf{(km)} & \textbf{Effective   Transmission} \linebreak \textbf{(nm)}                 & \textbf{Integration Time} \linebreak  \textbf{(s)} & \textbf{Observer} & \textbf{Country}

\\

\midrule

    2019-11-14T07:22.7 & 33.3 & 0.41 (N) & 589 & 750--1020 & 180 & Anthony Wesley & Australia \\ 
    2019-11-17T07:34.9 & 34.3 & 0.41 (N) & 589 & 1000--1020 & 180 & Anthony Wesley & Australia \\ 
    2019-11-18T07:53.2 & 34.6 & 0.41 (N) & 583 & 1000--1020 & 180 & Anthony Wesley & Australia \\ 
    2019-11-19T07:50.3 & 35.0 & 0.41 (N) & 583 & 1000--1020 & 180 & Anthony Wesley & Australia \\ 
    2019-11-20T07:50.5 & 35.3 & 0.41 (N) & 583 & 1000--1020 & 180 & Anthony Wesley & Australia \\ 
    2019-11-22T07:53.6 & 36.0 & 0.41 (N) & 578 & 1000--1020 & 180 & Anthony Wesley & Australia \\ 
    2019-11-23T08:13.9 & 36.3 & 0.41 (N) & 578 & 1000--1020 & 180 & Anthony Wesley & Australia \\ 
    2019-11-23T15:00.6 & 36.3 & 0.35 (S) & 663 & 884--900 & 300 & Emmanuel Kardasis & Greece \\ 
    2019-11-24T08:07.9 & 36.7 & 0.41 (N) & 573 & 1000--1020 & 180 & Anthony Wesley & Australia \\ 
    2019-11-27T14:19.0 & 37.7 & 0.35 (S) & 652 & 884--900 & 300 & Emmanuel Kardasis & Greece \\ 
    2019-11-29T07:58.2 & 38.4 & 0.41 (N) & 568 & 1000--1020 & 180 & Anthony Wesley & Australia \\ 
    2019-11-30T08:06.5 & 38.7 & 0.41 (N) & 563 & 1000--1020 & 180 & Anthony Wesley & Australia \\ 
    2019-12-01T14:41.0 & 39.1 & 0.35 (S) & 646 & 884--900 & 300 & Emmanuel Kardasis & Greece \\ 
    2019-12-02T08:09.8 & 39.4 & 0.41 (N) & 558 & 1000--1020 & 180 & Anthony Wesley & Australia \\ 
    2019-12-03T15:20.0 & 39.7 & 0.35 (S) & 641 & 884--900 & 300 & Emmanuel Kardasis & Greece \\ 
    2019-12-04T08:13.4 & 40.1 & 0.41 (N) & 558 & 1000--1020 & 180 & Anthony Wesley & Australia \\ 
    2019-12-05T08:02.4 & 40.4 & 0.41 (N) & 554 & 1000--1020 & 180 & Anthony Wesley & Australia \\ 
    2019-12-06T08:48.0 & 40.8 & 0.35 (S) & 635 & 850--1020 & 180 & Niall MacNeill & Australia \\ 
    2019-12-07T14:55.0 & 41.1 & 0.35 (S) & 630 & 884--900 & 300 & Emmanuel Kardasis & Greece \\ 
    2019-12-08T15:29.0 & 41.5 & 0.35 (S) & 630 & 884--900 & 300 & Emmanuel Kardasis & Greece \\ 
    2019-12-18T08:09.0 & 45.0 & 0.41 (N) & 527 & 1000--1020 & 180 & Anthony Wesley & Australia \\ 
    2019-12-18T15:23.0 & 45.0 & 0.35 (S) & 604 & 884--900 & 300 & Emmanuel Kardasis & Greece \\ 
    2020-01-11T11:34.0 & 53.7 & 0.50 (D) & 393 & 807--1050 & 249 & Tiziano Olivetti & Thailand \\ 
    2020-01-11T14:51.0 & 53.7 & 0.35 (S) & 547 & 884--900 & 300 & Emmanuel Kardasis & Greece \\ 
    2020-01-29T11:42.2 & 60.8 & 0.50 (D) & 356 & 807--1050 & 315 & Tiziano Olivetti & Thailand \\ 
    2020-02-01T11:27.6 & 62.0 & 0.50 (D) & 352 & 807--1050 & 415 & Tiziano Olivetti & Thailand \\

\bottomrule
\end{tabularx}
% \begin{adjustwidth}{+\extralength}{0cm}
% 		\noindent\footnotesize{Notes: (N) Newtonian Telescope, (S) Schmidt Cassegrain Telescope, (D) Dall-Kirkham Telescope. }
% 	\end{adjustwidth}
%\end{table}
\end{sidewaystable}

\begin{sidewaystable} \ContinuedFloat
%\begin{table}[H] \ContinuedFloat

\caption{\emph{Cont}.}
\begin{tabularx}{\textwidth}{cccccccc}
\toprule

\textbf{Date and Time (UT)}  & \textbf{Venus Phase} & \textbf{Telescope}  & \textbf{Best Spatial}  & \textbf{Effective} &  \textbf{Integration} & \textbf{Observer}  &
\textbf{Country} \\

\textbf{(YYYY-MM-DDThh:mm.m)} & \textbf{Angle (deg)} &  \textbf{Diameter (m)} & \textbf{Resolution (km)} & \textbf{Transmission (nm)} & \textbf{Time (s)} & & 

%\textbf{Date and Time (UT)} \linebreak \textbf{(YYYY-MM-DDThh:mm.m)} & \textbf{Venus Phase Angle} \linebreak \textbf{(deg)} & \textbf{Telescope Diameter} \linebreak \textbf{(m)} & \textbf{Best Spatial Resolution} \linebreak \textbf{(km)} & \textbf{Effective   Transmission} \linebreak \textbf{(nm)}                 & \textbf{Integration Time} \linebreak  \textbf{(s)} & \textbf{Observer}  & \textbf{Country}

\\
\midrule

    2020-02-01T15:51.0 & 62.0 & 0.35 (S) & 490 & 884--900 & 300 & Emmanuel Kardasis & Greece \\ 
    2020-02-05T11:38.2 & 63.7 & 0.50 (D) & 343 & 807--1050 & 415 & Tiziano Olivetti & Thailand \\ 
    2020-02-16T07:55.0 & 68.6 & 0.41 (N) & 387 & 1000--1020 & 180 & Anthony Wesley & Australia \\ 
    2020-02-24T17:27.0 & 72.3 & 0.35 (S) & 419 & 884--900 & 300 & Emmanuel Kardasis & Greece \\ 
    2020-02-24T18:48.5 & 72.3 & 0.35 (S) & 419 & 807--1050 & 380 & Paulo Casquinha & Portugal \\
    2020-02-28T17:41.7 & 74.2 & 0.28 (S) & 509 & 742--1050 & 240 & Antonio Gallardo & Spain \\ 
    2020-03-07T15:59.0 & 78.3 & 0.35 (S) & 377 & 850--1020 & 160 & Raimondo Sedrani & Italy \\ 
    2020-03-08T11:37.0 & 78.9 & 0.50 (D) & 269 & 807--1050 & na & Tiziano Olivetti & Thailand \\ 
    2020-03-09T11:32.0 & 79.4 & 0.50 (D) & 266 & 807--1050 & na & Tiziano Olivetti & Thailand \\ 
    2020-03-10T16:57.0 & 79.9 & 0.35 (S) & 367 & 850--1020 & 160 & Raimondo Sedrani & Italy \\ 
    2020-03-11T16:44.8 & 80.5 & 0.35 (S) & 366 & 884--900 & 300 & Emmanuel Kardasis & Greece \\ 
    2020-03-11T17:09.1 & 80.5 & 0.35 (S) & 366 & 1000--1050 & 300 & Luigi Morrone & Italy \\ 
    2020-03-11T16:30.0 & 80.5 & 0.35 (S) & 366 & 807--1050 & 60 & Joaquin Camarena & Spain \\ 
    2020-03-12T15:59.0 & 81.0 & 0.35 (S) & 360 & 807--1050 & 60 & Joaquin Camarena & Spain \\ 
    2020-03-13T11:22.0 & 81.6 & 0.50 (D) & 256 & 807--1050 & 482 & Tiziano Olivetti & Thailand \\ 
    2020-03-14T16:25.0 & 82.1 & 0.35 (S) & 354 & 807--1050 & 60 & Joaquin Camarena & Spain \\ 
    2020-03-15T15:10.0 & 82.7 & 0.40 (D) & 314 & 950--1050 & 180 & Yaroslav Naryzhniy & Ukraine \\ 
    2020-03-16T15:00.0 & 83.3 & 0.40 (D) & 313 & 950--1050 & 180 & Yaroslav Naryzhniy & Ukraine \\ 
    2020-03-16T16:01.0 & 83.3 & 0.28 (S) & 438 & 685--1050 & 1000 & Giovanni Calapai & Italy \\ 
    2020-03-16T16:54.4 & 83.3 & 0.35 (S) & 349 & 1000--1050 & 300 & Luigi Morrone & Italy \\ 
    2020-03-21T11:34.4 & 86.2 & 0.50 (D) & 237 & 807--1050 & 301 & Tiziano Olivetti & Thailand \\ 
    2020-03-21T17:35.0 & 86.2 & 0.25 (K) & 330 & 830--1020 & 205 & Dzmitry Kananovich & Estonia \\ 
    2020-03-21T17:10.2 & 86.2 & 0.35 (S) & 330 & 850--1020 & 160 & Raimondo Sedrani & Italy \\ 
    2020-03-21T17:35.0 & 86.2 & 0.35 (S) & 330 & 884--900 & 300 & Emmanuel Kardasis & Greece \\ 
    
\bottomrule
\end{tabularx}
% \begin{adjustwidth}{+\extralength}{0cm}
% 		\noindent\footnotesize{Notes: (N) Newtonian Telescope, (S) Schmidt Cassegrain Telescope, (D) Dall-Kirkham Telescope, (K) Klevtzov Cassegrain Telescope}
% 	\end{adjustwidth}
%\end{table}
\end{sidewaystable}

\begin{sidewaystable} \ContinuedFloat
%\begin{table}[H] \ContinuedFloat

\caption{\emph{Cont}.}
\begin{tabularx}{\textwidth}{cccccccc}
\toprule

\textbf{Date and Time (UT)}  & \textbf{Venus Phase} & \textbf{Telescope}  & \textbf{Best Spatial}  & \textbf{Effective} &  \textbf{Integration} & \textbf{Observer}  &
\textbf{Country} \\

\textbf{(YYYY-MM-DDThh:mm.m)} & \textbf{Angle (deg)} &  \textbf{Diameter (m)} & \textbf{Resolution (km)} & \textbf{Transmission (nm)} & \textbf{Time (s)} & & 

%\textbf{Date and Time (UT)} \linebreak \textbf{(YYYY-MM-DDThh:mm.m)} & \textbf{Venus Phase Angle} \linebreak \textbf{(deg)} & \textbf{Telescope Diameter} \linebreak \textbf{(m)} & \textbf{Best Spatial Resolution} \linebreak \textbf{(km)} & \textbf{Effective   Transmission} \linebreak \textbf{(nm)}                 & \textbf{Integration Time} \linebreak  \textbf{(s)} & \textbf{Observer}  & \textbf{Country}

\\
\midrule

    2020-03-26T15:20.0 & 89.3 & 0.40 (D) & 281 & 950--1050 & 180 & Yaroslav Naryzhniy & Ukraine \\ 
    2020-03-26T17:01.6 & 89.3 & 0.28 (S) & 394 & 742--1050 & 240 & Antonio Gallardo & Spain \\ 
    2020-03-31T16:15.0 & 92.6 & 0.40 (D) & 266 & 950--1050 & 180 & Yaroslav Naryzhniy & Ukraine \\ 
    2020-03-31T17:04.0 & 92.6 & 0.35 (S) & 296 & 884--900 & 300 & Emmanuel Kardasis & Greece \\ 
    2020-04-05T16:40.0 & 96.1 & 0.40 (D) & 250 & 950--1050 & 180 & Yaroslav Naryzhniy & Ukraine \\ 
    2020-04-10T16:18.0 & 99.9 & 0.35 (S) & 261 & 884--900 & 300 & Emmanuel Kardasis & Greece \\ 
    2020-04-10T17:00.1 & 99.9 & 0.35 (S) & 261 & 1000--1050 & 300 & Luigi Morrone & Italy \\ 
    2020-04-20T11:31.7 & 108.5 & 0.50 (D) & 187 & 807--1050 & 247 & Tiziano Olivetti & Thailand \\ 
    2020-04-25T17:27.0 & 113.4 & 0.35 (S) & 211 & 884--900 & 300 & Emmanuel Kardasis & Greece \\
    \midrule 
    2020-03-08T11:37.6 & 78.9 & 0.50 (D) & -- & 320--380 & 1029 & Tiziano Olivetti & Thailand \\ 
    2020-03-09T11:39.7 & 79.4 & 0.50 (D) & -- & 320--380 & 804 & Tiziano Olivetti & Thailand \\ 
    2020-03-10T17:04.0 & 79.9 & 0.35 (S) & -- & 320--380 & 300 & Emmanuel Kardasis & Greece \\ 
    2020-03-11T16:55.0 & 80.5 & 0.35 (S) & -- & 320--380 & 300 & Emmanuel Kardasis & Greece \\ 
    2020-03-13T11:31.3 & 81.6 & 0.50 (D) & -- & 320--380 & 569 & Tiziano Olivetti & Thailand \\ 
    2020-03-14T16:32.2 & 82.1 & 0.25 (M) & -- & 320--380 & 500 & Christian Viladrich & France \\ 
    2020-03-15T15:10.0 & 82.7 & 0.40 (D) & -- & 350--410 & 180 & Yaroslav Naryzhniy & Ukraine \\ 
    2020-03-16T15:00.0 & 83.3 & 0.40 (D) & -- & 350--410 & 180 & Yaroslav Naryzhniy & Ukraine \\    
\bottomrule
\end{tabularx}
Notes:\\
(N) Newtonian Telescope,\\ 
(S) Schmidt Cassegrain Telescope,\\
(D) Dall-Kirkham Telescope,\\
(K) Klevtzov Cassegrain Telescope.

%\end{table}
\end{sidewaystable}
	
%\finishlandscape

%\begin{paracol}
%\linenumbers
%%\switchcolumn

The images from amateur observations were obtained thanks to the technique of {\textit{lucky imaging}}, where thousands of images are captured in a few minutes with fast planetary cameras and then stacked to obtain a single one more nitid and with better resolution {(by up to a factor of four in poor seeing conditions)  \cite{Farsiu2004, Law2006}}.  Software such as Autostakkert ~({\url{http://www.autostakkert.com} , accessed on 2021) or Registax ~({\url{ http://astronomie.be/registax}}, } accessed on 2021) \citep{Berrevoets2012} were used following this strategy: (a) selection of the highest quality images, (b) alignment, (c)~stacking, and (d) processing of the final image to get a sharper version \cite{Kardasis2016}. To obtain these observations, sensitive CMOS cameras (e.g., ZWO290 mm) were equipped on $\sim$0.2--0.5 m aperture telescopes. In the case of NIR images, Venus's clouds exhibit lower contrast than in UV images, and a different image processing is required \citep{Peralta2019GRL}.

{The first step in the analysis was to properly define the navigation of each image, i.e., the correct orientation (north--south), fitting of the limb, and the phase of the planet to a synthetic outline (calculated by software's ephemeris).} {For this, and further processing of the images, we used} the free software WinJupos ~({\url{http://www.grischa-hahn.homepage.t-online.de}} , accessed on 2021) \citep{Hahn2012}. We also used this software to measure the motions of clouds patterns (latitude--longitude position, size, and wind speed). The phase and the  ephemeris of the planet are automatically calculated by the software. Two rotation periods are implemented in WinJupos for the determination of the planet longitudes: System 1 (S1) accounting for the surface rotation with a period of $\sim$243 days, and System 2 (S2) with a period of 4.2 days, corresponding to the superrotating upper clouds in UV images \citep{Sanchez-Lavega2017}. For creating maps of the dayside middle clouds with NIR images, we used a slower rotation rate with a period of almost 5 days (calculated from the average equatorial speed of CD events, see Table~\ref{table:CD-properties}). CD speeds were measured using feature tracking of the following wave-front edge (FWFE henceforth) in pairs of images (with time separation ranging from a few hours to some~days).

Most of the cloud motions were measured near the equator, where the CD is customarily found and also because the spatial resolution approaches the best that can be attained with ground-based telescopes (see Table \ref{table:NIR-observations}). {Using the standard deviations of the measurements as a metric, we compared those derived from the optimal fitting of the images with those obtained by slightly changing the parameters that affect the navigation (orientation and limb fitting), by adjusting to the largest and lowest acceptable values (by visual inspection). We found that limb fitting (with an error of 2--3 pixels) was the one resulting in systematically larger standard deviations than the orientation. However, even in those cases, the standard deviations were always smaller than those produced by the image resolution.} 

We obtained an estimate of the speed error by two methods. In the first approach, we used the theoretical resolution of each telescope to calculate the possible error (\mbox{$\delta \upsilon=\delta x / \Delta t$}, where $\delta x$ corresponds to the theoretical expected image resolution according to the telescope's aperture expressed in meters, and $\Delta t$ is the time separation between image pairs). Image pairs separated by a temporal interval of $\sim$2--8 h allow us to measure wind speeds with errors in the range $\sim$10--40\mps. Image pairs separated by $\Delta t \sim$ 5--10 days provide smaller errors in the range of $\sim$1 \mps, because of the large time difference. We  note here that these error estimates are lower limits, as they consider optimal seeing conditions. However, given the lucky imaging approach and deconvolution techniques, this can be {lowered} ever more. A second approach was to measure wind speed from various latitude bins of 5--12.5$^{\circ}$. The standard deviation of these measurements within each bin results in errors of $\sim$1 \mps for $\Delta t \sim$ 5--10 days and $\sim$5--25 \mps for $\Delta t \sim$ 2--8 h, respectively. Given the similarity between the two approaches, we opted to quote the error estimates as derived from the second method (i.e., the statistical approach). Although this approach depends on the determination of the CD by ``visual inspection'', it takes into account the high quality of the selected images. %case where the discontinuity was recognized in observations separated by some days and in some cases in the same day separated by some hours).

%\end{paracol}

% Table with CD properties
%\input{Table_CD_Properties}

%\startlandscape

\begin{sidewaystable}
%\begin{table}[H] 
%    \small
    \caption{Summary of cloud discontinuity events' properties as derived from this work. The first two columns present the pair of images that were used to extract the wind measurements near the equator. For each detection, we show the orientation (column 3), the maximum latitude range (column 4), the maximum length (column~5), the width at the equator (column 6), and the derived rotation period (column 7) and zonal speed (column 8). }
%    \small
    \label{table:CD-properties}   
    \begin{tabularx}{\textwidth}{cccccccc}
    \toprule 
 %   Date,Time (1st obs.)$^1$ & Date,Time (2nd obs.)$^1$ & Orientation$^2$  & Max. Latitude Range & Max. Length$^3$  & Width at equator$^4$ & Rotation period & Zonal speed$^5$ \\     
    
     \textbf{Date and Time (UT) $^1$} & \textbf{Date and Time (UT) $^1$ }& \textbf{Orientation $^2$}  & \textbf{Maximum} & \textbf{Maximum}  & \textbf{Width at} & \textbf{Rotation}  & \textbf{Zonal} \\
     \textbf{(1st obs.)} & \textbf{(2nd obs.)} &  & \textbf{Latitude Range} & \textbf{Length$^3$} & \textbf{Equator $^4$}  &  \textbf{Period}  & \textbf{Speed $^5$ }\\ 
    \textbf{(YYYY-MM-DDThh:mm)} & \textbf{(YYYY-MM-DDThh:mm)} & \textbf{(degrees)}  & \textbf{(degrees}) & \boldmath{($10^3\,\textrm{km}$)}  & \boldmath{($10^3\,\textrm{km}$)} & \textbf{(Earth days)}  &  \textbf{(m\boldmath{$\cdot$ s$^{-1}$})} \\ 
    \midrule 

  2019-11-23T08:14* & 2019-11-23T15:51 & {124} & 20$^{\circ}$ N--9$^{\circ}$ S & 3.3 & 0.6 & 4.1 $\pm$ 0.9 & 105 $\pm$ 19 \\ 
  2019-11-23T15:51 & 2019-12-03T15:20* & {97} & 26$^{\circ}$ N--28$^{\circ}$ S & 5.7 & 0.6 & 5.06 $\pm$ 0.01 & 84.9 $\pm$ 0.1  \\ 
  2019-12-03T15:20 & 2019-12-08T15:29* & {39} & 14$^{\circ}$ N-- 17$^{\circ}$ S & 5.1 & 1.1 & 5.06 $\pm$ 0.01 & 84.9 $\pm$ 0.1  \\ 
  2019-12-18T08:09* & 2019-12-18T15:23 & 77 & 27$^{\circ}$ N--28$^{\circ}$ S & 5.9 & 0.6 & 4.2 $\pm$ 0.2 & 104 $\pm$ 5  \\ 
  2020-01-11T11:34* & 2020-01-11T14:51 & {137} & 12$^{\circ}$ N--14$^{\circ}$ S & 3.7 & 0.5 & 4.3 $\pm$ 0.8 & 100 $\pm$ 16  \\ 
  2020-02-01T11:27 & 2020-02-01T15:51* & {39} & 19$^{\circ}$ N--25$^{\circ}$ S & 6.5 & 0.6 & 4.0 $\pm$ 0.4 & 108 $\pm$ 9  \\ 
  2020-02-16T07:55* & -- & {84} & 13$^{\circ}$ N--25$^{\circ}$ S & 4.0 & 0.3 & -- & --  \\ 
  2020-02-24T17:27 & 2020-02-24T18:36* & {85} & 28$^{\circ}$ N--28$^{\circ}$ S & 6.1 & 0.5 & 3.5 $\pm$ 0.9 & 125 $\pm$ 26  \\ 
  2020-02-24T18:36 & 2020-02-28T17:41* & {66} & 15$^{\circ}$ N--13$^{\circ}$ S & 3.3 & 0.6 & 4.18 $\pm$ 0.02 & 103.4 $\pm$ 0.5  \\ 
  2020-02-28T17:41 & 2020-03-11T16:45* & {100} & 25$^{\circ}$ N--30$^{\circ}$ S & 5.9 & 0.5 & 6.00 $\pm$ 0.01 & 71.6 $\pm$ 0.1  \\ 
  2020-03-11T16:45 & 2020-03-16T15:00* & {95} & 23$^{\circ}$ N--36$^{\circ}$ S & 6.3 & 0.7 & 5.14 $\pm$ 0.02 & 83.9 $\pm$ 0.3  \\ 
  2020-03-16T15:00 & 2020-03-21T17:35* & {88} & 27$^{\circ}$ N--32$^{\circ}$ S & 6.3 & 0.5 & 5.10 $\pm$ 0.01 & 84.5 $\pm$ 0.2  \\ 
  2020-03-21T11:34 & 2020-03-21T17:35 & -- & -- & -- & -- & 5.09 $\pm$ 0.08 & 84.7 $\pm$ 1.3  \\ 
  2020-03-21T17:35 & 2020-03-26T15:20* & {94} & 13$^{\circ}$ N--33$^{\circ}$ S & 5.0 & 0.4 & 5.10 $\pm$ 0.02 & 84.6 $\pm$ 0.4  \\ 
  2020-03-26T15:20 & 2020-03-31T17:04* & {99} & 20$^{\circ}$ N--29$^{\circ}$ S & 5.2 & 1.0 & 5.15 $\pm$ 0.03 & 83.6 $\pm$ 0.5  \\ 
  2020-03-31T17:04 & 2020-04-05T16:40* & {90} & 31$^{\circ}$ N--25$^{\circ}$ S & 6.0 & 0.5 & 4.91 $\pm$ 0.03 & 87.8 $\pm$ 0.5  \\ 
  2020-04-05T16:40 & 2020-04-10T16:12* & {93} & 31$^{\circ}$ N--29$^{\circ}$ S & 6.5 & 0.6 & 5.09 $\pm$ 0.01 & 84.7 $\pm$ 0.2  \\ 
  2020-04-10T16:12 & 2020-04-20T11:32* & {80} & 10$^{\circ}$ N--20$^{\circ}$ S & 3.3 & 0.3 & 5.05 $\pm$ 0.01 & 85.4 $\pm$ 0.1  \\ 
   &  &  &  &  &  &  &   \\ 
  \multicolumn{2}{c}{AVERAGE}  & 88 & 20.8$^{\circ}$ N--21.5$^{\circ}$ S & 5.2 & 0.6 & 4.8 $\pm$ 0.2 & 92.1 $\pm$ 4.6  \\    

    \bottomrule 
  
  \end{tabularx} 
  
Notes:\\
$^1$ We note with an * the observations from which the properties of CD event were measured.\\
$^2$ Degrees relative to equator (scale 0--180$^{\circ}$) right to left.\\
$^3$ North--south direction, error of $\pm 0.2 \times 10^3\,\textrm{km}$.\\
$^4$ East--west direction, error of $\pm 0.2 \times 10^3\,km$.\\
$^5$ Near equator, i.e., within 5$^{\circ}$ N--5$^{\circ}$ S.\\
%\\Notes: \\
%$^1$ We note with an * the observations from which the properties of CD event were measured. \\
%$^2$ Degrees relative to equator (scale 0$^{\circ}$-180$^{\circ}$) right to left.\\
%$^3$ North-South direction, error of $\pm 0.2 \times 10^3\,\textrm{km}$.\\
%$^4$ East-West direction, error of $\pm 0.2 \times 10^3\,km$.\\
%$^5$ Near equator, i.e., within 5$^{\circ}$ N - 5$^{\circ}$ S.

%\end{table}
\end{sidewaystable}

%\finishlandscape

The position and the inclination of the CD were also measured using the FWFE. Inclinations were measured with respect to the equator (0--180$^{\circ}$). {Maps were created with equirectangular projection that simply displays the relative position of large formations}. The gaps correspond to areas where no observations were performed or where the spatial resolution was very poor. Latitudes higher then 60$^{\circ}$ were also excluded because of the low resolution and severe distortion near the polar region expected in equirectangular~projections.

%Amateur digital observations are based on a technique referred to as ``lucky imaging'' \citep{Farsiu2004, Law2006} \comment{}{Do we add Law2006 and Farsiu 2004?}. Thousands of images are captured in a few minutes with fast planetary cameras. Software such as Autostakkert\footnote{\url{http:// www.autostakkert.com}} or Registax\footnote{\url{ http://astronomie.be/registax}} \citep{Berrevoets2012} is then used to select the best images, align, stack and sharpen them \citep{Kardasis2016}. They use sensitive CMOS cameras (like the ZWO290mm) combined with telescopes of $\sim0.2-0.5\,m$ apertures. The UV observations are made with filters in the range of $\sim320-380\,nm$. The NIR observations are made with filters in the range $0.685-1020\,nm$ where an extra sharpening is required due to the low contrast of Venusian features in that wavelength range. The resolution is $200-400\, km$ depending on telescope aperture, observing wavelength, observing conditions and apparent disk diameter.

%At the epoch of the main CD event (March 21, 2020) Venus presented solar elongation of 46$^{\circ}$ East, 23 arcsec in diameter, and 52\% of illuminated disk. 

\subsection{Venus Images Taken with JAXA's \akatsuki}

Amateur observations during March 2020 were compared with nearly simultaneous observations from \akatsuki \citep{Murakami2019,Nakamura2016}. This gave us the chance to test whether the CD events can be visualized at other vertical levels and if there is any interaction with other planetary scale waves at the upper clouds. {Such phenomena include} the Y-feature, {a dark cloud structure observed for decades in UV images that looks like a horizontally oriented Y letter visible every $\sim$4 days} \citep{Peralta2015,Imai2019}{, and} the stationary bow-shape gravity waves \citep{Kouyama2017,Peralta2017NatAstro}. These phenomena at the upper clouds can be observed in images taken by the Longwave Infrared Camera (LIR) \citep{Fukuhara2011} and the Ultraviolet Imager (UVI) \citep{Yamazaki2018}) onboard \akatsuki. The bolometer LIR maps the brightness temperature of the upper clouds within $\sim$60--75  km~\citep{Taguchi2007} in the spectral band 8--12 $\mu$m, both on the dayside and nightside, while the camera UVI senses the albedo of the cloud tops at $\sim$70  km \citep{Yamazaki2018,Horinouchi2018} using filters centered at 283 and 365 nm. Five events of the CD during March {(11 March 2020, 16 March  2020, 21 March  2020, 26 March  2020, and 31 March 2020)} were studied in combination with \akatsuki images, which covered orbits 142--144 of the mission \citep{Murakami2017UVIl2,Murakami2018UVIl3,Murakami2017LIRl2,Murakami2018LIRl3}. During March 2020, the phase angle of Venus as observed from \akatsuki varied between 30$^{\circ}$ and 160$^{\circ}$, while the spatial resolution of UVI and LIR images at lower latitudes ranged 12--90 and 40--360 km/pixel, respectively. The UVI images were photometrically corrected, {i.e., correcting for the limb-darkening effect} \citep{Lee2017}. Both UVI and LIR images were processed and projected as described \mbox{by \citet{Goncalves2020}}. A devoted software written in IDL was used to process and perform measurements in the \akatsuki images \citep{Peralta2018ApJS}.

\section{Results and Discussion}
\label{s:results}
\begin{figure*}
%\centering
\includegraphics[width=0.9\textwidth]{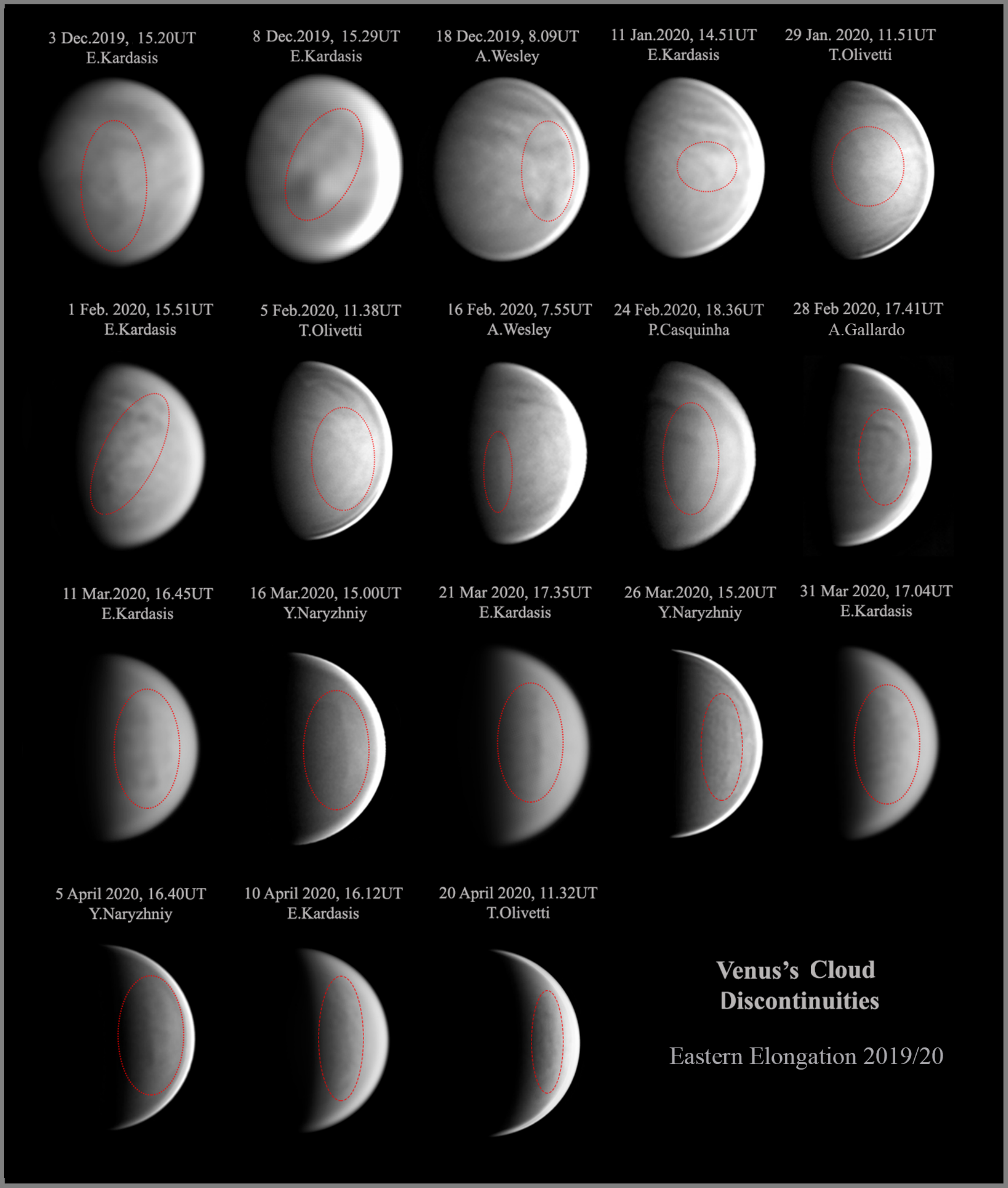}
\caption{Observations of cloud discontinuities, observed in the 2019/2020 eastern elongation of Venus, showing different morphologies. Details of the observations are presented in Table~\ref{table:NIR-observations}.}
\label{f:Orig-Observs}
\end{figure*}

\begin{figure*}
%\centering
\includegraphics[width=0.9\textwidth]{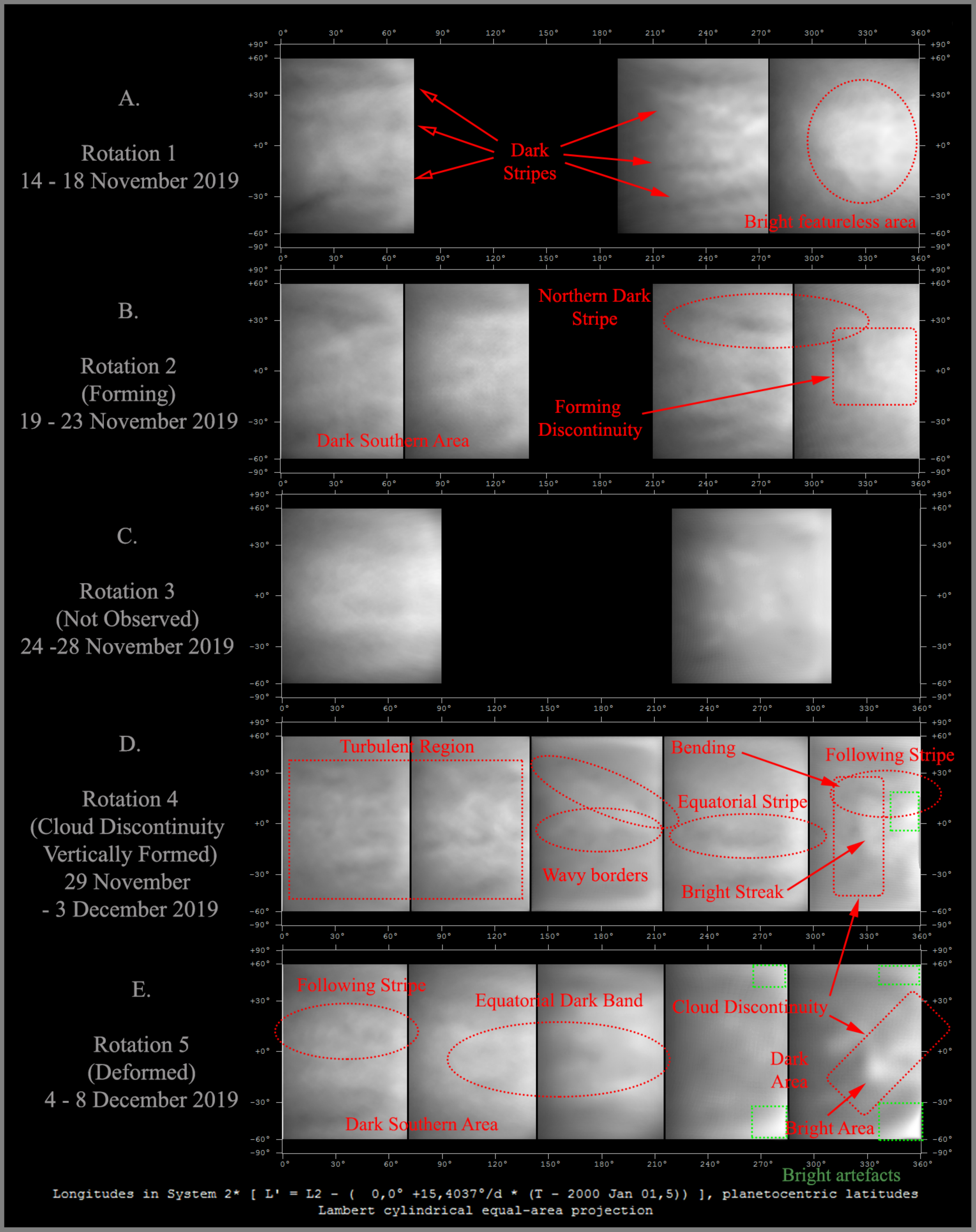}
\caption{Time composites of the of the middle clouds (NIR $\sim$900--1000 nm) of Venus showing large atmospheric features when the cloud discontinuity is observed. These large-scale features are marked with red ovals. Equirectangular projections (60$^{\circ}$ N to 60$^{\circ}$ S) placed from left to right with increasing dates: (\textbf{A}) {14--18 November 2019} (A. Wesley, E. Kardasis), \mbox{(\textbf{B}) {19--23 November 2019}} (A. Wesley, E. Kardasis), (\textbf{C}) {24--28 November 2019} (A. Wesley, E. Kardasis), (\textbf{D}) {29 November--3 December 2019} (A. Wesley, E. Kardasis, N.  MacNeill), (\textbf{E}) {4--8 December 2019} (A. Wesley, E.Kardasis). Details of the observations are presented in Table~\ref{table:NIR-observations}.}
\label{f:composites}
\end{figure*}

Three years after its last detection on the dayside middle clouds in December 2016~\citep{Peralta2020}, the CD was captured again in November--December of 2019 (see Figures~\ref{f:Orig-Observs} and \ref{f:composites}). Unfortunately, from mid-December 2019 until early March 2020, we have gaps in the observations that prevented a continuous monitoring of these new events of the CD. From March 2020, the presence of a CD became more evident, manifesting as a long dark feature sometimes followed by a brighter streak (see Figure~\ref{f:Orig-Observs}). The larger number of observations enabled us to monitor with more detail this phenomenon until the middle of April, when the CD seemed to~vanish.

%
%\begin{paracol}
%\linenumbers

%\begin{figure*}
%\centering
%\includegraphics[width=\textwidth]{CDE_Nonember-December2019-Comments_MKardasis.pdf}
%\caption{Time composites of the  of the middle clouds (NIR $\sim$900--1000 nm) of Venus showing large atmospheric features before, during and after the Cloud Discontinuity. These large-scale features are marked with red ovals. Equirectangular projections (60$^{\circ}$ N to 60$^{\circ}$ S and 0.36$^{\circ}$ per pixel) placed from left to right with increasing dates: (A) 14--18 November (A. Wesley, E. Kardasis), (B) 19--23 November (A. Wesley, E. Kardasis), (C) 24--28 November (A. Wesley, E. Kardasis) (D) 29 November--3 December (A. Wesley, E. Kardasis, N. McNiall) (E) 4--8 December (A. Wesley, E. Kardasis). Panel (F) Observations of cloud discontinuities, observed in 2019/2020 eastern elongation of Venus, showing different morphologies.}
%\label{f:cdeNovDec2019}
%\end{figure*}

% Image with original ground-based observations
\unskip

\subsection{CD during November--December 2019}
The first clear manifestation of CD occurred in late November {2019} (Figures~\ref{f:Orig-Observs} and \ref{f:composites}). On {18 November 2019}, Venus displayed no features on the area where we would expect to observe the CD. There are three dark stripes or bands (Northern, Equatorial, Southern) in the proceeding area. On {23 November 2019}, the CD seems apparent as an extension of the Northern Dark Stripe, which was previously reported as a cloud pattern frequently preceding the arrival of the discontinuity on the dayside middle clouds during the year 2016 \citep{Peralta2020} (see {Figure~1b}). On {3 December 2019}, the CD is observed between 26$^{\circ}$ N and 28$^{\circ}$ S, bending at its northern end and eastward extending $\sim$5700 km as an horizontal sharp dark stripe similar to those reported by \citet{Peralta2019Icarus} on the nightside lower clouds. On the west of CD, an equatorial stripe was formed also. In the central image of Figure~\ref{f:composites}D, a bright equatorial area is observed, surrounded north and south with wavy borders.

Taking advantage of the smaller phase angle for these earlier observations, we studied the morphology and development of CD along several rotations through maps of time composites (see Figure \ref{f:composites}) similar to the ones created by \citet{Peralta2020} for the nightside lower clouds. Due to the difficulty in accurately measuring the middle clouds' mean flow with amateur images, the equirectangular projections of the NIR images were combined after shifting them according to the average CD speed ($\sim$5 day period, \mbox{see \citet{Peralta2020})} at the equator rather the mean wind speed from passive tracers. During the last rotation \mbox{(Figure~\ref{f:composites}E)}, the CD changed its inclination (starting with a tilt of {124}$^{\circ}$ relative to the parallels, then {97}$^{\circ}$ and finally {39}$^{\circ}$) and a bright area follows the CD in the southward of the equator. The CD drifted to the west with an average equatorial zonal speed of $-94.7\pm 6$\mps, with a derived period of $4.6\pm 0.3$ days.

\subsection{CD during January--April 2020}
From November--December 2019 to March 2020, the increasing phase angle and more sparse data coverage hindered the monitoring of the CD, though some of the images show its presence (see Figure~\ref{f:Orig-Observs}). During this period, the CD suffered noticeable morphological changes after each full revolution, which is consistent with the observations by \akatsuki /IR1 during the year 2016 \citep{Peralta2020}. Between January 2020 and April 2020, the CD was always observed between \mbox{31$^{\circ}$ N and 36$^{\circ}$ S}, with a length within 3300--6500 km and tilt relative to the parallels ranging {39}--{137}$^{\circ}$. Its average speed was $-91\pm 4$\mps with a period of $4.8\pm 0.2$ days.

On {11 March} {2020}, our observations exhibited an intensification of the CD, being apparent as a long dark sharp discontinuity followed by a brighter cloudy area, and barely suffering distortions on its shape and size along {11, 16, and 21 March 2020}. A similar event was also reported in August 2016 \citep{Peralta2020}. Combining images from several ground-based observers, we were able to track the CD until {25 April 2020}, and we estimated for the CD a rotation period of about 5 days. From {26 March 2020} onwards, it appears to have shrunk, partly dissipated, and displayed much lower contrast during the next weeks. In general, it spanned between 31$^{\circ}$ N and 36$^{\circ}$ S, ranging 3300--6500 km in length. The dark streak was sometimes followed by a bright one. The CD average width at the equator was $\sim$600 km, {with a lower value of $\sim$300 km (similar to the spatial resolution of the images) and up to $\sim$ 1100 km}. The March--April event presented some constant pro\-per\-ties such as inclination (almost vertical relative to the equator, {80}--{100}$^{\circ}$) and speed ($-$83.6 to $-$87.8\mps, with an average of $84.9\pm0.4$ \mps, and a period $5.08\pm 0.03$ days.).

\begin{figure}[t]
\centering
\includegraphics[width=0.45\textwidth]{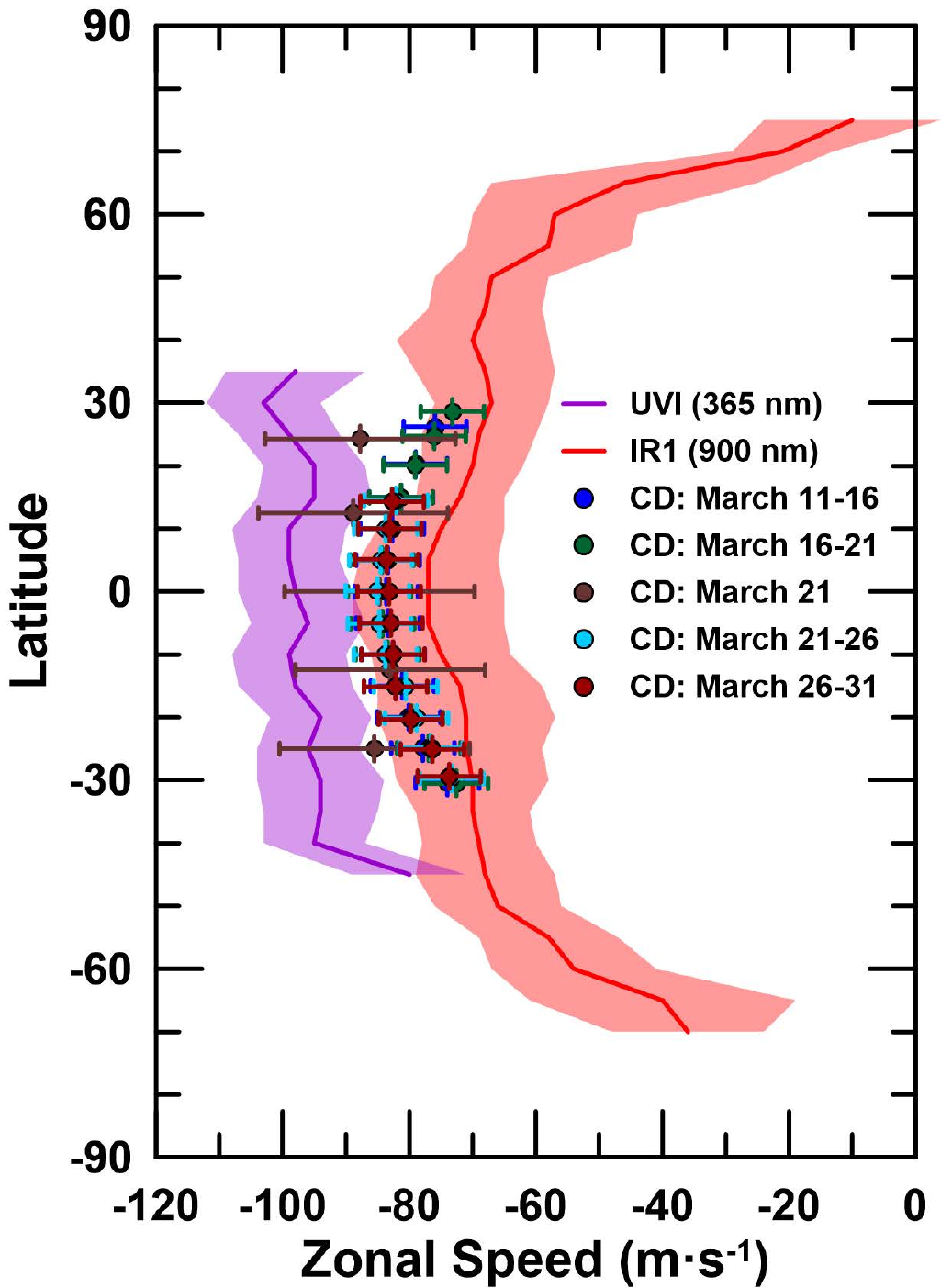}
\caption{Zonal speeds of the clouds' discontinuity (CD) during March 2020 (dots) compared with the zonally-averaged profiles of the zonal wind for the top of the upper clouds at $\sim$70 km (violet line) and the middle clouds within 50--55 km \citep{Peralta2019GRL} (red line). The speeds of the discontinuity and the winds at the cloud tops were obtained from ground-based near-infrared images and \akatsuki/UVI 365 nm images respectively, considering observations matching the CD events: {11, 16, 21, 26, and 31 March 2020}. For the winds at the middle clouds, we considered the profile obtained \mbox{by \citet{Peralta2019GRL}} during 2016. As a result of the worse spatial resolution at higher latitudes, speed vectors had larger error bars (errors of about $\pm15$\mps for $\Delta$t$\sim$6 hours and $\pm5$\mps for $\Delta$t$\sim$5 terrestrial days).}\label{f:PhaseSpeed-vs-Winds}
\end{figure}

Figure~\ref{f:PhaseSpeed-vs-Winds} exhibits the zonal speeds of the discontinuity during the CD events of {11, 16, 21, 26, and 31  March 2020} apparent in images of Venus taken by ground-based observers (see Figure~\ref{f:Orig-Observs}). These zonal speeds are compared with the zonal winds measured with 365 nm images taken by \akatsuki/UVI during the same dates as the CD events occurring during {11, 16, 21, and 26 March 2020}. In the case of the winds at the middle clouds, it was not possible to identify reliable passive tracers in the NIR ground-based images due to their low contrast, and the profile of winds obtained by \akatsuki/IR1 in 2016 \citep{Peralta2019GRL} is shown instead. The CD speeds are consistent with the manifestation of a Kelvin wave \citep{Peralta2020}, with phase speeds faster than the background winds at the middle clouds observed in the 900~nm images from \akatsuki/IR1. On the other hand, the phase speeds of the CD are slower than the winds measured in the UVI 365 nm images sensing the top of the upper clouds, supporting the argument that the CD cannot be observed at the clouds' top because it becomes dissipated during its vertical propagation \citep{Peralta2020}.

% FIGURE WITH CD SPEEDS COMPARED WITH WINDS

\subsection{Long-Term Evolution}
\begin{figure}[h]
\centering
\includegraphics[width=0.8\columnwidth]{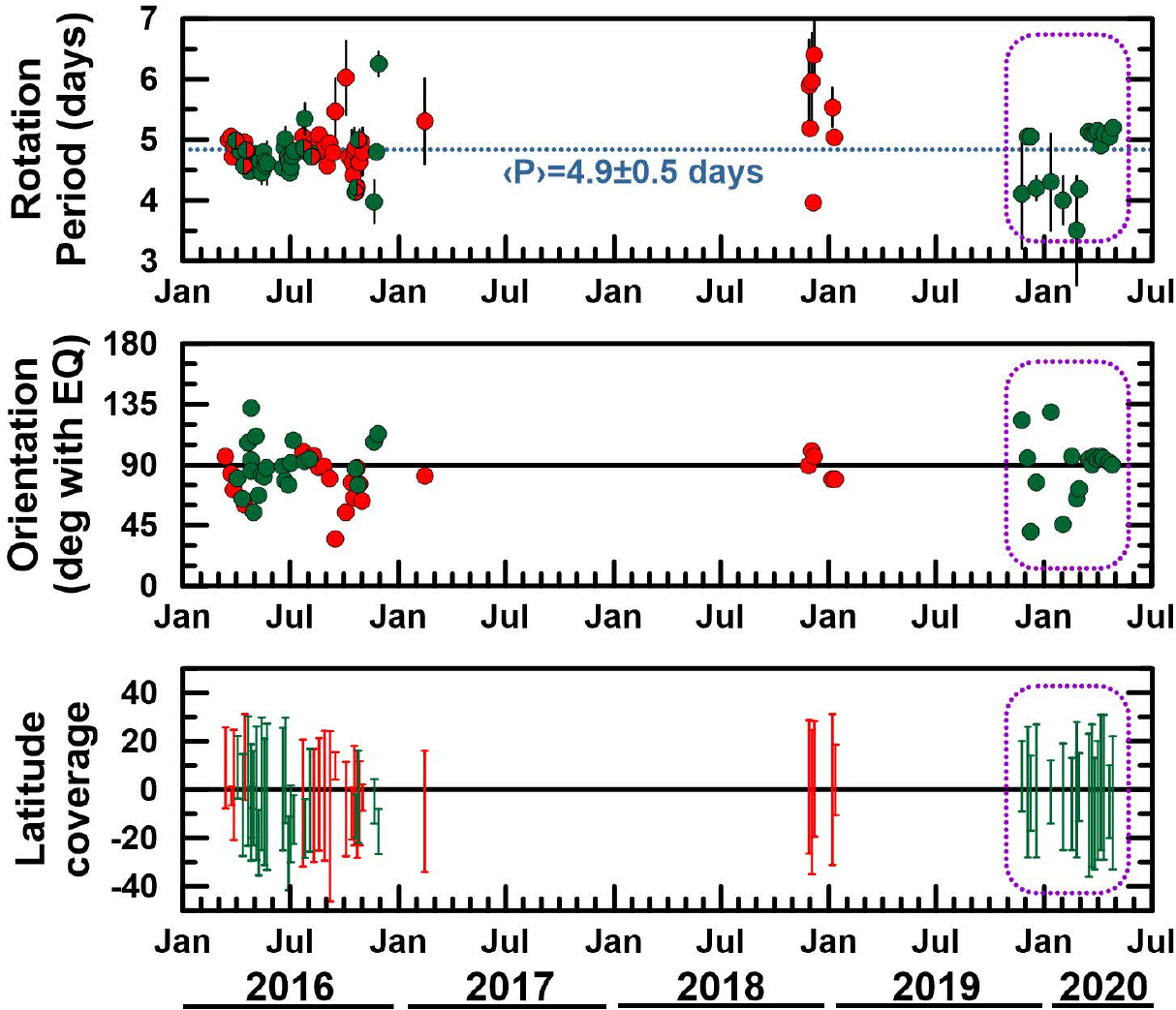}
\caption{Long-term evolution of CD properties. Combination of past (2016--2018) with current (2019--2020) measurements (marked with the purple outline). The rotation period, orientation, and latitude coverage of the disruption along 2016--2020 are exhibited. The periods were measured from the position of the disruption at the equator in images separated by hours and several days. When the CD did not intersect the equator, we considered its longitude closest to the equator. The mean period is shown with a blue dotted line. Day and night-side data are shown in green and red, respectively.}
\label{f:cd_evolution}
\end{figure}

Figure~\ref{f:cd_evolution} exhibits the rotation period, orientation, and latitude coverage of the discontinuity during the years 2016--2020 from Venus images taken by the orbiter \akatsuki \citep{Peralta2020} and ground-based observations by NASA's Infrared Telescope Facility \citep{Peralta2020} and small telescopes (this work). Concerning the rotation period of the CD, we obtain a period of $4.8\pm 0.2$ days considering the averaged zonal speed ($92.1\pm 4.6$\mps) from the 17 CD events documented in this work (see Table~\ref{table:CD-properties}), while the mean rotation period considering the full data from 2016 to 2020 can be updated to $4.9\pm 0.5$ days, the same as estimated by \citet{Peralta2020}. The discontinuity exhibited noticeable variations in its orientation between November 2019 and February 2020, with a clear tendency to keep an orientation perpendicular to the equator from March until April 2020. {Given that gravity waves are subject to refraction as they propagate, due to the varying background atmospheric temperature structure and winds~\citep{Heale2015}, these changes in the orientation of the discontinuity might be tentatively attributed to a slight atmospheric refraction whose evidence should be investigated in future works}. Finally, the CD events reported in this work confirm the hemispherical asymmetry for the discontinuity previously reported by \citet{Peralta2020}.  

% FIGURE WITH CD PROPERTIES ALONG 2016-2020

\subsection{The Cloud Tops during the CD Events of March 2020}

The Y-feature is the most noticeable planetary-scale pattern that can be identified in the ultraviolet images that sense the top of the upper clouds at the dayside of Venus \citep{Boyer1961,Belton1976a,Rossow1980,Kouyama2012}. The Y-feature shares some of the characteristics of the lower clouds' discontinuity: the Y-feature is a recurrent phenomenon subject to cycles of generation and destruction \citep{Rossow1980}, it manifests about the equator, it propagates faster than the mean zonal flow \citep{Kouyama2012}, and it has been long interpreted as the manifestation of a kelvin-type wave \citep{Peralta2015}. On the other hand, the Y-feature exhibits tends to have a rotation period of about 4 days \citep{Imai2019} (generally faster than the discontinuity, see Figure~\ref{f:cd_evolution}), it can extend to higher latitudes \citep{Rossow1980,Peralta2015}, and it has never been reported on observations of altitudes below the top of the clouds \citep{Belton1991,Peralta2019GRL}, which is in agreement with the idea that the Y-feature may be vertically trapped \citep{Peralta2015}. 

\begin{figure*}
\centering
\includegraphics[width=0.95\textwidth]{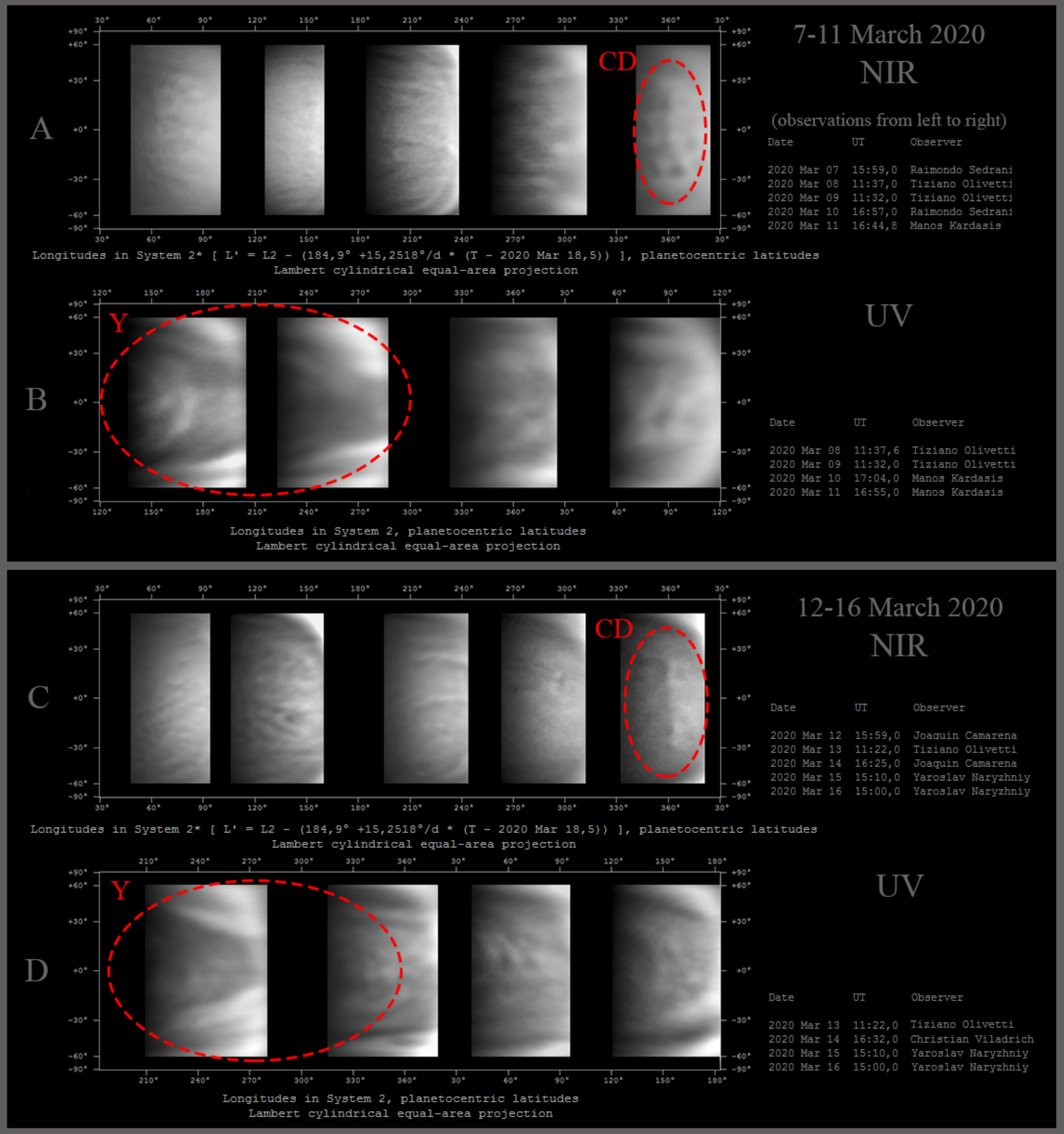}
\caption{Comparison among simultaneous NIR amateur maps showing the CD  position in the days observed in relation to Y-feature. NIR map between {7--11 March 2020 (\textbf{A}) and 12--16 March 2020 (\textbf{C})} from amateur observations  compared with UV amateur maps presenting  the Y-feature {on 8 March 2020 to 11 March 2020 (\textbf{B}) and 13 March 2020 to 16 March 2020 (\textbf{D})}. NIR and UV maps were constructed by taking as  reference images the simultaneous observations in both wavelengths when the CD was observed.}
\label{f:compareUVI-NIR}
\end{figure*}

\begin{figure}
\centering
\includegraphics[width=0.7\textwidth]{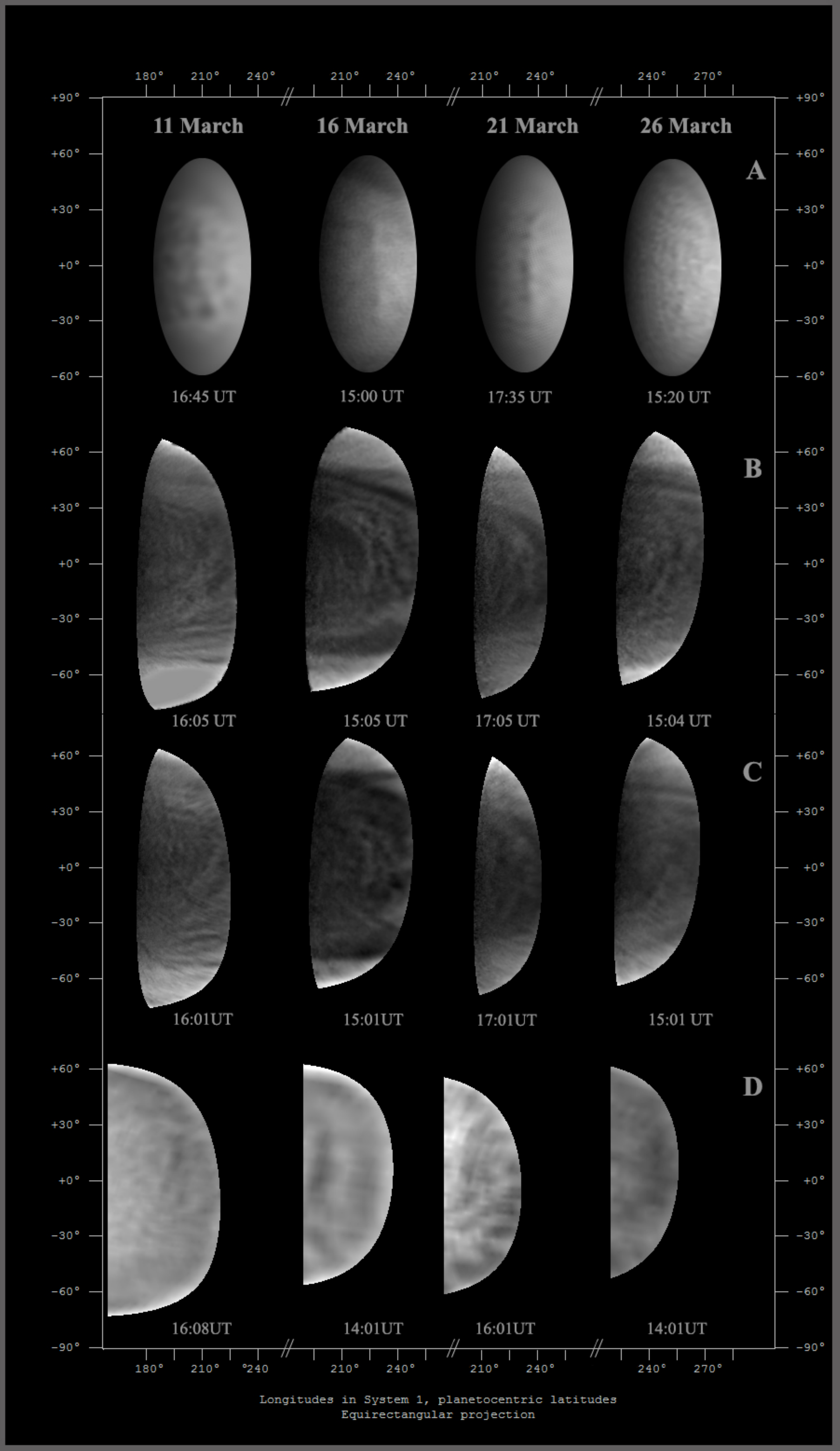}
\caption{Contemporaneous views of NIR amateur set ($\sim$700-1000 nm) where the cloud discontinuity is observed (\textbf{A}), along with \akatsuki /UVI at 365 nm (\textbf{B}), UVI at 285 nm (\textbf{C}), and LIR at 10 $\mu$m (\textbf{D})~where the cloud discontinuity seems absent.}
\label{f:compareAkatsuki}
\end{figure}

Figures~\ref{f:compareUVI-NIR} and \ref{f:compareAkatsuki} display a comparison between the morphology of the cloud tops (UV images) and the middle clouds (NIR images) during some of the CD events occurring in March 2020. The cycle of the discontinuity at the dayside middle clouds is confronted by that of the Y-feature at the cloud tops in Figure \ref{f:compareUVI-NIR}, using NIR and UV images obtained by ground-based observers and time composites similar to those published \mbox{by \citet{Sanchez-Lavega2016}}. When comparing the cycles of {7--11 March and 12--16 March 2020}, we observe that the position of the sharp vertical streak-form of the CD (evident in NIR images) does not seem linked to the propagation of the Y-feature in UV images or to any specific cloud pattern in its cycle. The time composites also display an evident phase-lag between both phenomena, supporting that we may be observing independent waves that propagate at different vertical levels.

%\begin{figure*}[H]

\akatsuki/UVI imaging of the top of the clouds provide higher resolution to study the details of individual cloud features and patterns that cannot be resolved by ground-based telescopes. In fact, UVI allows observing Venus with two filters: 283 nm to sense the absorption of the SO$_2$ and 365~nm to observe the absorption due to the unknown absorber~\citep{Yamazaki2018}. Figure~\ref{f:compareAkatsuki} displays a representative sampling of upper clouds in images obtained very close to the time of the NIR amateur data and covers the exact area of CD observations on March 2020. %Please ensure that your intended meaning has been retained. 
The sharp vertical form of CD (Figure~\ref{f:compareAkatsuki}A) due to variations of the middle-cloud optical thickness is also absent in high-quality UV data (Figures~\ref{f:compareAkatsuki}B,C), confirming previous findings by \citet{Peralta2020} and supporting the idea that the critical level where the CD dissipates should be located between the middle clouds and upper clouds' top (see Figure~\ref{f:PhaseSpeed-vs-Winds}).

\subsection{Thermal Emission from the Upper Clouds during the CD Events of March 2020}

Large stationary gravity waves appear mostly above four specific highland regions in the low latitudes, specially when these regions are in the local afternoon \citep{Kouyama2017}. The exact mechanism for generating these stationary waves at the cloud top is as yet unclear, although there is a general agreement that we must be visualizing Lee waves triggered at the surface of Venus {(stationary waves in the atmosphere, happening where winds encounter obstacles such as high mountains; }\cite{Peralta2017NatAstro,Fukuhara2017NatGeo,Navarro2018,Kitahara2019,Lefevre2020}). These planetary scale waves tend to be aligned almost north--south with a slight curvature, and they present silhouette similarities with the cloud discontinuity.

During March 2020, \akatsuki/LIR exhibits clear stationary waves in the brightness temperature images taken during {11--16 and 21 of March 2020}. The wavefronts were located above Alta Regio and almost aligned in a north--south direction spanning several thousand kilometers. They were not visible during {26 and 31 March} {2020}. In Figure~\ref{f:compareAkatsuki}D, the discontinuity is not apparent in LIR images acquired in almost the same time as the CD events, while the LIR image taken {on 26 March 2020} does not cover the region where the CD was observed at the middle clouds with ground-based observations (Figure~\ref{f:compareAkatsuki}A). {Since the weighting function of LIR extends within about 50--60 km above the surface \citep{Taguchi2007}, LIR might be (partially) sensitive to the thermal emission from middle clouds (50.5--56.5 km) as well. In this sense, the absence of the CD in the LIR images may provide new hints to better constrain the altitude range for the critical level of the Kelvin wave interpreted to be responsible for the CD.}

\section{Summary}
\label{s:summary}
In this work, we report the reappearance of the equatorial discontinuity at the dayside middle clouds of Venus \citep{Peralta2020}, which is a global-scale atmospheric wave comparable with other planetary phenomena such as the Y-feature~\citep{Peralta2015} and the giant stationary bow-shape wave~\citep{Fukuhara2017NatGeo}. We studied its morphology, the evolution, and the properties of the most prominent events in 2019/2020 eastern elongation of Venus, combining both NIR and UV data from ground-based amateur observations and \akatsuki imaging.

The most intense event occurred in March 2020, which was visible in NIR images as a dark vertical streak followed by a bright streak. In general, the CD events extended approximately 20$^{\circ}$ above and below the equator, displaying a variable length (\mbox{3300--6500 km}) with {widths in the range of $\sim$300--1100 km (with an average of 600 km)}. Although the orientation displayed noticeable variations, it is generally perpendicular to the equatorial plane. In total, we have measured the mean rotational period from all CD events to be $\sim$4.9 days with a mean wind speed of $\sim$92.1 \mps, which is consistent with \citet{Peralta2020}. We also confirm that {the morphology of the CD is often hemispherically asymmetric} \citep{Peralta2020}.
%We acquired size, position, shape and speed measurements of the phenomenon. 
%In general it was placed between 31$^{\circ}$ N - 36$^{\circ}$ S, with a variable length of 3300 to 6500 km and a mean width of 600 km. Its mean orientation relative to the equatorial plane is 88$^{\circ}$ but ranges from 40$^{\circ}$ to 123$^{\circ}$. 

In general, the speeds we obtained are higher/lower when compared with the average zonal speeds of the middle/upper clouds found by \akatsuki in 2016/2020. In combination with imaging, we conclude that the CD is a phenomenon limited to the middle clouds of the atmosphere of Venus. When combining ground-based amateur images with \akatsuki UV and LIR imaging, we find no direct relation between the CD in the middle and the phenomena in upper clouds (such as Y-feature or the bow-shape waves). Moreover, the evident phase lags between these phenomena point to a scenario with independent waves, which propagate at different vertical levels. Our results further support that the CD dissipates during its vertical propagation, with the critical level located between the middle and upper clouds.

With the current work, we aim to both contribute to the additional characterization of this phenomenon and to highlight the importance of the amateur contributions. Amateur observations can play an important role especially when used complementary to professional ones. Moreover, systematic monitoring by amateurs can trigger, and of course, support coordinated observing campaigns, such as the next campaigns of observations coordinated with \akatsuki~({\url{http://pvol2.ehu.eus/bc/Venus/}}) of Venus flybys of NASA's \textit{PARKER} \citep{Wood2022} on 21  August 2023 and 6 November 2024. The professional--amateur collaboration offers an invaluable resource in the advance of our understanding of the atmosphere of~Venus.

%%%%%%%%%%%%%%%%%%%%%%%%%%%%%%%%%%%%%%%%%%%%%%%%%%%%%%%%%%%%%%%%
%  ACKNOWLEDGMENTS
%%%%%%%%%%%%%%%%%%%%%%%%%%%%%%%%%%%%%%%%%%%%%%%%%%%%%%%%%%%%%%%%

\section*{Funding}
J.P. thanks EMERGIA funding from Junta de Andaluc\'{i}a Spain (code: EMERGIA20\_00414). G.M. acknowledges funding support from the European Research Council (ERC) under the European Union's Horizon 2020 research and innovation programme (Grant agreement No. 772086). M.I. is supported by the Japan Society for the Promotion of Science (JSPS) (Grant-in-Aid for JSPS Research Fellow: JP21J00351).

\section*{Acknowledgments}
We are grateful for support from all the members of the Akatsuki mission. We are also grateful to many amateur observers that observed Venus intensively during the investigated period  providing data for this research. Special thanks to the observer Raimondo Sedrani for using some of his observations. E.K. would like to offer his special thanks to Grischa Hahn developer of WinJupos software for providing help and advice for the proper use. Moreover, E.K. would like to acknowledge the continuous support of his wife Dimitra.

\section*{Data availability statement}
This study was based on ground-based images of Venus provided by worldwide amateur observers and Akatsuki observational data provided by ISAS. Ground-based observations are publicly available in the Venus section of ALPO-Japan at \url{http://alpo-j.sakura.ne.jp/Latest/Venus.htm} (accessed on 2021). All original Akatsuki/UVI  and LIR data are available via Data ARchives and Transmission System (DARTS), provided by the Center for Science-satellite Operation and Data Archive (C-SODA) at ISAS/JAXA, and NASA PDS Atmospheres Node. The analyzed data are the L2b and L3bx products, and these data are provided in the UVI  datasets (accessed on 2020)
 \url{https://data.darts.isas.jaxa.jp/pub/pds3/vco-v-uvi-3-cdr-v1.0/} and at \url{https://data.darts.isas.jaxa.jp/pub/pds3/extras/vco_uvi_l3_v1.0/}, and LIR datasets  at \url{https://data.darts.isas.jaxa.jp/pub/pds3/vco-v-lir-3-cdr-v1.0/} and \url{https://data.darts.isas.jaxa.jp/pub/pds3/extras/vco_lir_l3_v1.0/}.

%% ------------------------------------------------------------------------ %%
%% References and Citations
%% ------------------------------------------------------------------------ %%
%\bibliography{references}

%\end{adjustwidth}

\end{document}